# Ultrafast high-harmonic nanoscopy of magnetization dynamics

Sergey Zayko[1], Ofer Kfir[1,2], Michael Heigl[3], Michael Lohmann[1],
Murat Sivis[1,2], Manfred Albrecht[3], and Claus Ropers[1,2]

[1] 4th Physical Institute – Solids and Nanostructures, University of Göttingen, 37077 Göttingen, Germany.
[2] Max Planck Institute for Biophysical Chemistry, 37077 Göttingen, Germany.
[3] Institute of Physics, University of Augsburg, Universitätsstrasse 1, 86159 Augsburg, Germany

Light-induced magnetization changes, such as all-optical switching, skyrmion nucleation, and intersite spin transfer, unfold on temporal and spatial scales down to femtoseconds and nanometers, respectively. Pump-probe spectroscopy and diffraction studies indicate that spatio-temporal dynamics may drastically affect the non-equilibrium magnetic evolution. Yet, direct real-space magnetic imaging on the relevant timescale has remained challenging. Here, we demonstrate ultrafast high-harmonic nanoscopy employing circularly polarized high-harmonic radiation for real-space imaging of femtosecond magnetization dynamics. We observe the reversible and irreversible evolution of nanoscale spin textures following femtosecond laser excitation. Specifically, we map quenched magnetic domains and localized spin structures in Co/Pd multilayers with a sub-wavelength spatial resolution down to 16 nm, and strobosocopically trace the local magnetization dynamics with 40 fs temporal resolution. Our approach enables the highest spatio-temporal resolution of magneto-optical imaging to date. Facilitating ultrafast imaging with an extreme sensitivity to various microscopic degrees of freedom expressed in chiral and linear dichroism, we envisage a wide range of applications spanning magnetism, phase transitions, and carrier dynamics.

Localized magnetic textures are essential elements of hard-drives, magnetic random-access memories [1], novel storage schemes [2–4], and potential building blocks for next-generation logic devices [4–7]. Skyrmions and domain walls, in particular, attract considerable attention as topological excitations that can be nucleated, erased and controllably translated, equivalent to the writing and register-shifting of logical bits [3,4,7,8]. Due to the inherently localized nature of such magnetic features, their coupling with the environment is governed by nanoscale heterogeneity, while exchange and spin-orbit interactions define the intrinsic femtosecond time scale of magnetic dynamics. Whereas ultrafast demagnetization has been studied for decades [9], recent experiments based on extreme-UV and soft-X-ray scattering suggest a coupling between spatial magnetic properties and the temporal scale of the demagnetization process [10,11]. Similarly, magneto-transport phenomena involve an ultrafast spatio-temporal response, such as superdiffusive spin-currents [12,13], hot electrons [14], or the spin Seebeck effect [15,16], ranging down to a few tens of femtoseconds [17–19].

Access to combined nm-fs spatio-temporal resolution, as required for real-space observations of spin dynamics at the fundamental limits, has remained a grand experimental challenge. Time-resolved measurement schemes were developed to access the magnetic dynamics with sub-100 nm spatial and picosecond temporal resolution. These include implementations of spin-polarized scanning tunneling microscopy [20], Lorentz-contrast electron microscopy [21–27], scanning transmission x-ray microscopy [28,29], high-resolution ptychography [30,31] and tomography [32,33], as well as full-field magneto-optical imaging approaches including Fourier transform holography (FTH) [29,34,35] and coherent diffractive imaging (CDI) [36,37]. Magneto-optical Kerr microscopy [38] and photoemission electron microscopy [39], on the other hand, offer femtosecond snapshots of magnetic textures, albeit at somewhat lower spatial resolution. To date, the 2014 pioneering work by von Korff-Schmising and co-workers [40] remains the only experimental demonstration of ultrafast magnetization dynamics imaged with sub-100-nm spatial and 100-fs-temporal resolution, using FTH at a free-electron laser facility. Thus, exploring the frontiers of ultrafast magnetism still requires more widely accessible experimental techniques capable of simultaneous nanometer spatial and femtosecond temporal resolutions.

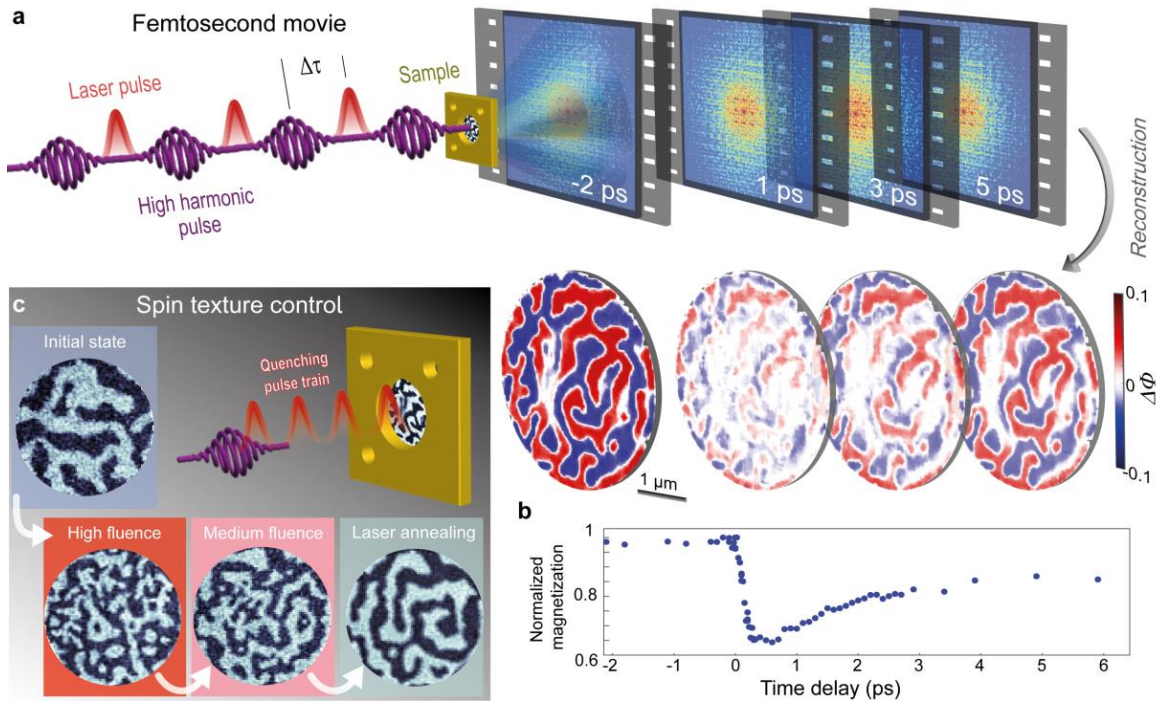

Fig. 1. Ultrafast high-harmonic nanoscopy. **a** A magnetic sample is excited with a femtosecond laser pulse and probed with a circularly polarized high-harmonic beam (wavelength of 21 nm) at a variable delay Δτ. For each time delay between the pump and probe pulses, a quantitative real-space image of the magneto-optical phase is reconstructed from the diffraction pattern of the high-harmonic beam. **b** Plot of the spatially-averaged normalized magnetization within the field of view as a function of delay. **c** *In-situ* observation of laser-induced control of the magnetization state. Images show the resulting patterns when the sample is quenched after excitation at high/medium fluence, and after laser annealing. Note in particular the highly localized structures with diameters less than 100 nm.

In this work, we introduce ultrafast dichroic nanoscopy based on high-harmonic radiation, reaching unprecedented spatio-temporal resolution in full-field magneto-optical imaging. Exploiting the X-ray magnetic circular dichroism (XMCD) of cobalt in Co/Pd multilayer structures, we trace the ultrafast response of nanoscale magnetic textures to femtosecond laser excitation. We capture laser-induced spin textures, ranging from micrometer-scale domains to ~80-nm-sized magnetic bubbles, with a sub-wavelength spatial resolution down to 16 nm. Ultrafast movies of nanoscale spin dynamics are recorded with a temporal resolution of 40 fs.

The experimental principle is depicted in Fig. 1a. Femtosecond pulses from a Ti:Sapphire laser amplifier (central wavelength of 800 nm, pulse energy up to 3.5 mJ, 35 fs pulse duration) are split into a pump and a probe arm. The pump pulses optically excite the sample, and circularly polarized radiation form high-harmonic generation (HHG) in a gas cell serves as the femtosecond probe for magnetic imaging. The 38th harmonic order (wavelength of ~21 nm), which provides XMCD contrast near the M-edge of cobalt, is selected and focused onto the sample by a toroidal grating monochromator, whereas the remaining harmonic orders are blocked. The diffraction pattern of the radiation transmitted through the sample is collected by a charge-coupled device (CCD) camera placed a few centimeters downstream. The positioning of the camera determines the collected scattering angle and the theoretical (diffraction) limit of the spatial resolution. Real-space images are retrieved algorithmically by an iterative phasing of the recorded far-field intensities, based on a combination of FTH and coherent diffractive imaging (CDI). Guiding the CDI algorithm with FTH drastically improves the image recovery [41–43] and allows us to significantly increase the spatial frequencies up to which the phases are consistently reconstructed [41]. Both phase and absorption contrast can be achieved via the choice of probe wavelength near the absorption edge [44,45]. Imaging of reversible magnetization dynamics is conducted stroboscopically (Fig. 1a) at 1 kHz repetition rate, whereas irreversible changes of the magnetic texture (Fig. 1c) are mapped after switching off the pump beam.

The magnetic sample is a multilayer structure of cobalt and palladium (Co/Pd) deposited on membranes made of silicon-nitride ($Si_3N_4$) or silicon. The backside of the sample is coated with gold (180-200 nm thick) which is opaque for the high-harmonic radiation. Fields of view (FOV) with diameters of 3 to 6 μm are formed by removing the gold using focused ion beam (FIB) etching. Additionally, an array of holes is milled through the entire sample thickness including the substrate and the magnetic film (c.f. Fig. 2a). The hole-array is designed such that the strong auxiliary field emanating from these holes covers the entire detector and interferometrically enhances the weak magneto-optical scattering signal from the FOV aperture. This signal-enhancement scheme drastically improves the dynamic range and reduces the required dose by more than one order of magnitude compared to a conventional FTH or CDI approach. Two of the holes (diameters of 200-400 nm) are well-separated to allow for a direct recovery of low-resolution holographic information that facilitates subsequent high-resolution iterative phase retrieval.

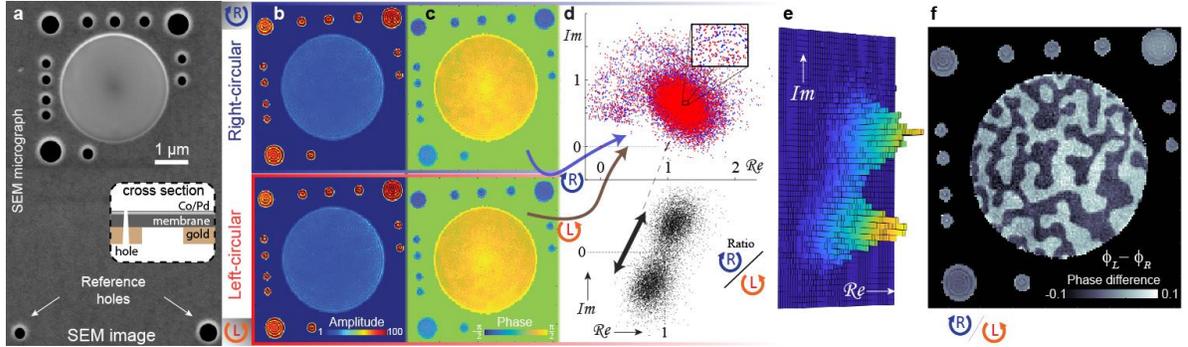

Fig. 2. Dichroic imaging employing coherent signal enhancement. **a** Scanning electron microscop (SEM) image of the sample showing a circular 3.1 µm field of view (FOV) , auxiliary holes for signal enhancement and two distant reference holes (bottom) for Fourier-transform holography (see methods). Inset: Sample cross section. **b, c** Exit-field magnitude and phase obtained with left- (L) and right-handed (R) circularly polarized high-harmonic beams. **d** Scatter plot of pixel-by-pixel complex exit-wave amplitudes in the FOV (top) and of their ratio, i.e., the dichroic signal (bottom). **e** 2D-histogram of pixels with a given complex dichroic signal. **f** Dichroic phase contrast image exhibiting domains with out-of-plane magnetization.

Extracting pure magnetic information requires the joint phase retrieval (phasing) of the far-field intensities recorded with left- and right-handed circularly polarized illumination. An image recorded with a single helicity is dominated by non-magnetic contrast, characterized by a higher magnitude of the transmission (Fig. 2b) and a phase offset (Fig. 2c) in the holes compared to the film within the circular FOV. Only weak variations in the FOV are observed for single-helicity reconstructions, including fringes near the perimeter from wave-guiding and edge-diffraction effects [46,47]. Some contrast arises from thickness variations and contaminations of the specimen and the substrate. Illustrating the importance of a dichroic reconstruction, Fig. 2d shows the complex amplitudes of all pixels in the single-helicity exit waves within the circular FOV, exhibiting considerable scatter in amplitude and phase. The dichroic signal, however, obtained by a pixel-by-pixel ratio of the complex amplitudes (L/R), shows a very well-defined double-lobed histogram distribution, reflecting the quasi-binary contrast in the final dichroic image (phase contrast in Fig. 2f).

We have found this dichroic microscopy approach to be very robust and have recorded hundreds of magnetization maps for different samples, material systems, masks, and numerical apertures. Figure 3 shows a set of dichroic phase-contrast images of magnetic textures with half-pitch spatial resolutions between 27 and 16 nm. Note that these are direct reconstructions without filtering or pixel interpolation. We consistently reach a diffraction-limited single-pixel resolution in each of the scattering geometries used. In real space, the diffraction-limited resolution is evident in sharp contrast changes at domain walls that span a single pixel. In a far-field analysis, the resolution is evaluated using the phase retrieval transfer function (PRTF) [48,49] – a standard measure for the reliability of phase retrieval procedures. The PRTF values, as shown in Fig 3c, stay above the 0.5 mark, indicating confidence in the retrieved phases even at the edges and the corners of the CCD sensor, corresponding to spatial frequencies (half-pitch resolutions) of 16 nm and 12 nm, respectively.

A high consistency and robustness of the demonstrated approach allows for *in-situ* imaging with only few-second delays between the raw data acquisition and the subsequent image reconstruction, providing for real-time feedback on experimental parameters. We have utilized this feature in the laser-manipulation of domain structures shown in Fig. 1c. In these experiments, a high laser excitation fluence was chosen to demagnetize the sample. Abrupt interruption of the excitation by blocking the laser beam leads to quench of the magnetic film. We have found that the resulting spin textures drastically depend on the excitation parameters [27], ranging from large micrometer-scale domains over highly localized bubbles to rather fractures textures. Further elucidation of this optical domain control will be a subject of future work, as we will now focus on the ultrafast capabilities of high-harmonic nanoscopy.

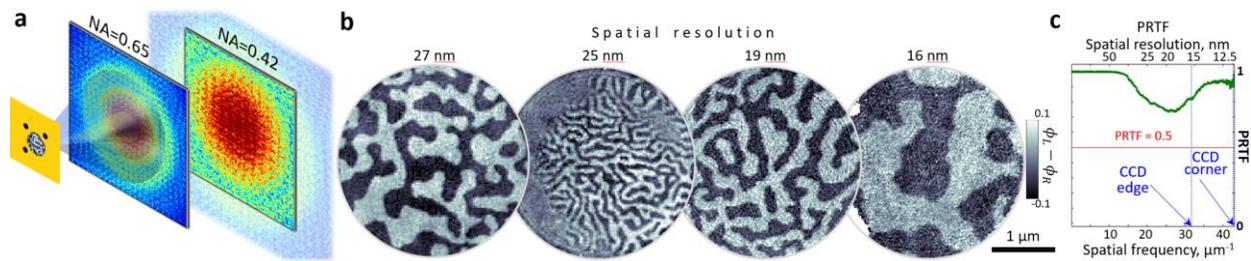

Fig. 3 High-resolution full-field magnetic imaging. **a** Illustration of the diffraction data collection at various numerical apertures (NA). **b** Dichroic phase contrast reconstructions for the diffraction data recorded with 21 nm wavelength at NA=0.39, 0.42, 0.55 and 0.65 corresponding to a half-pitch spatial resolution of 27 nm, 25 nm, 19 nm, and 16 nm, respectively. Raw experimental reconstructions without image processing, averaging or interpolation. **c** The phase-retrieval transfer function (PRTF) computed for the diffraction data recorded at NA=0.65. The PRTF plot indicates high consistency of the retrieved phases up to the edges and corners of the CCD sensor, corresponding to spatial resolutions of 16 nm and 12 nm, respectively.

Imaging with high-harmonic radiation inherently allows for time-resolved studies on the femtosecond to attosecond time scale [17,50,51]. In the following, we utilize this capability to record ultrafast magnetization movies of the multilayer films using 35-fs laser pump and sub-10 fs high-harmonic probe pulses (see Supplementary Materials). Figures 4a and b show exemplary XMCD phase-contrast frames of the spin dynamics movies for two different samples and fluences, reconstructed with a spatial resolution of 40 nm. The pump fluences were chosen to induce considerable demagnetization while minimizing irreversible changes of the domain pattern. Comparing consecutive frames (cf. Figs. 4a and b), the most pronounced observation is the overall reduction of the magnetization immediately after the optical pump ($\Delta\tau = 0$) with a well-resolved timescale of 200 fs, and a subsequent partial recovery over a few picoseconds. Figures 4c and d show high-fidelity reconstructions of a domain pattern before the pump and near the maximum suppression of the magnetization around 1 ps. The change of the normalized magnetization is displayed in Fig. 4e, revealing a standing-wave-like pattern with features separated by about half the optical wavelength, and suggesting considerable scattering of the pump beam at the mask edges. In a local subwavelength hotspot in the right of the FOV, a suppression down to about 20% is observed. Segmentations in areas of different local fluence (Fig. 4g) yield de- and remagnetization dynamics that strongly vary across the FOV (Fig. 4h), and which, accordingly, also differ in shape from the spatially-averaged dynamics evaluated for different fluences (Fig. 4f).

A major part of the spatially-varying magnetization dynamics observed can be attributed to the structured excitation, most likely including the slower recovery at regions of higher local fluence (Fig. 4h) [17,52]. Despite the high spatiotemporal resolution of the approach, up to now, we have not found unambiguous evidence for a significant softening of domain walls as may have been expected from the characteristics of superdiffusive spin currents and previous diffraction experiments [3,11,53–57]. Several reasons may account for this observation. First, the spatial resolution achieved may not yet be sufficient to resolve a broadening if it only involves a change by a small fraction of the native domain wall width. Secondly, stroboscopic imaging relies on reversible changes of the domain pattern. We observe that irreversible changes to the real-space domain structure already occur at fluences for which the local magnetization is suppressed by about 50 %. Diffraction experiments, however, suggest that considerable domain wall-softening may require higher fluences. Further studies, possibly with more controlled local excitation and higher spatial gradients, may be necessary to clearly identify such spatiotemporal magnetization dynamics and map nanoscale spin currents in real space.

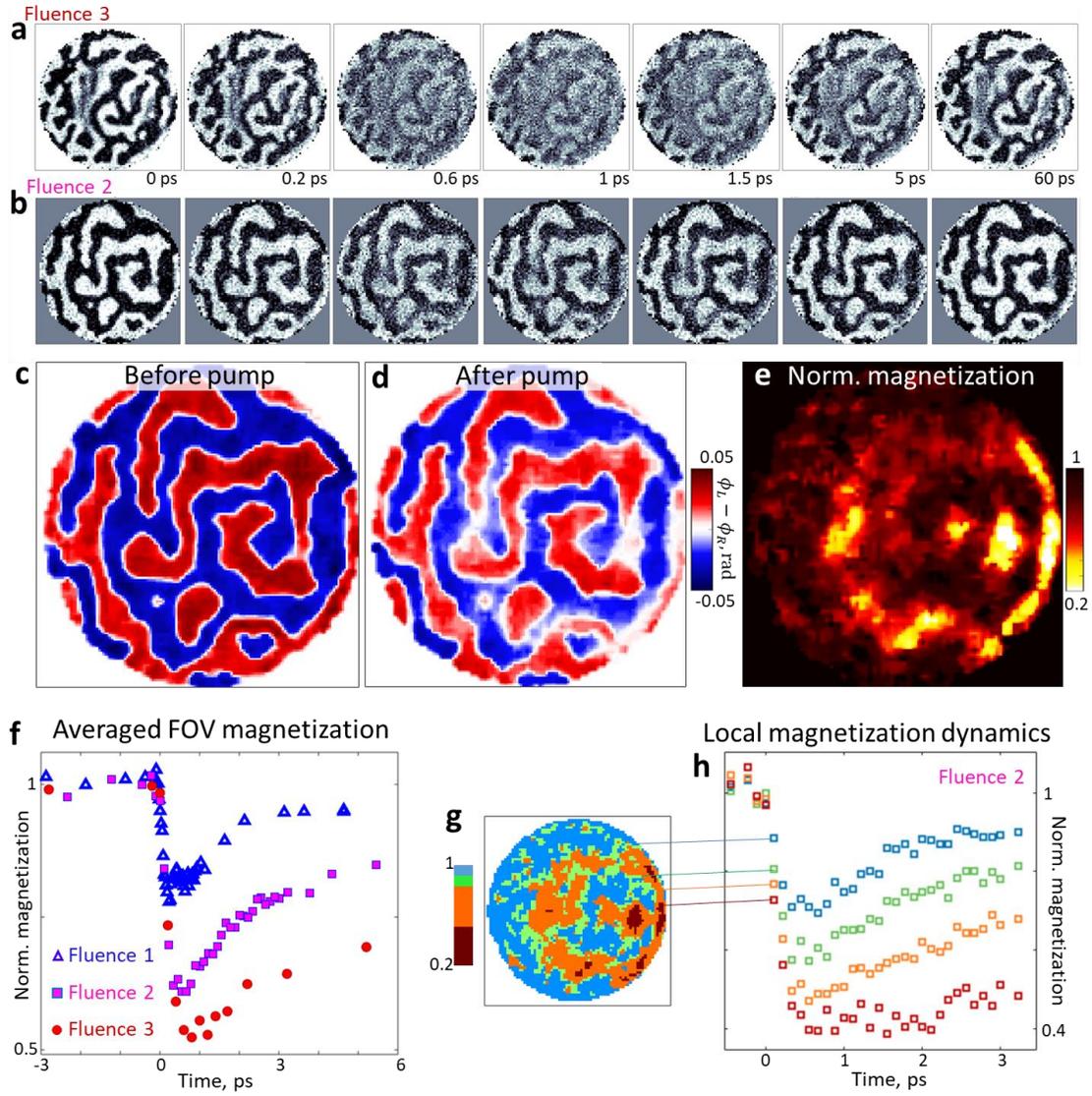

Fig. 4. Ultrafast high-harmonic magnetic nanoscopy. **a, b** Exemplary frames from movies of magnetization dynamics for two samples driven at different fluences. Nominal incident fluences: 1.4(1) mJ/cm² (a) 1.3(1) mJ/cm² (b). **c** Interpolated averaged images over 4 frames before the optical pumping (data from panel b). **d** Image after pumping (4 frames around 1 ps). Note the strong local variations in the suppression of the magnetization. **e** The demagnetization map is the ratio between the magnetic-contrast before and apfter the pump highlighting the regions with stronger demagnetization. **f** Evolution of the average magnetization (absolute value) within the field of view, for different laser pump fluences. **h** Local magnetization dynamics for 4 different regions of the sample highlighted in panel **g**.

In conclusion, we have developed ultrafast high-harmonic dichroic nanoscopy for robust full-field imaging of magnetic textures and their femtosecond evolution on the nanoscale. These results represent both the highest spatial and the highest spatiotemporal resolution in full-field magneto-optical imaging obtained to date, irrespective of the wavelength, technique or the radiation source used. Reaching the diffraction limit with a high-coherence source allows for such results using a wavelength one to two orders of magnitude longer than typically required to access sub-50 nm resolutions. We envisage a wide range of applications for high-harmonic dichroic nanoscopy using linear or circular dichroism, with many further examples in ultrafast and element-specific spintronic imaging, but also extending to the tracking of hot carrier populations, structural and electronic phase transitions, and chemical transformations.

## Methods

**HHG source.**
High-harmonic up-conversion is driven with a bi-circular two-color field tailored using a MAZEL-TOV apparatus from a linearly polarized laser beam with the central wavelength at 800 nm. We optimize the generation conditions to improve the harmonic yield near the maximum of the magneto-optical phase signal of cobalt M-edge [44,45]. The generated harmonic spectrum is dispersed with a toroidal grating and the 38th harmonic (~21 nm wavelength) is isolated with a slit before it enters the imaging chamber. We use a diffraction grating to remove adjacent harmonics and ensure sufficient temporal coherence across the mask for high-resolution lensless imaging [58,59]. Moreover, the spatial coherence from the bichromatic circular driver is higher compared to a conventional HHG scheme with linearly polarized excitation due to a reduced number of allowed electron trajectories [60,61]. The observation of a full suppression of every 3rd harmonic order is a robust and reliable indication for generation conditions corresponding to circularly polarized harmonics. However, in order to maximize the magneto-optical signal, the circular polarization has to be provided at the sample position. This can be achieved by adjusting the waveplate angle of the MAZEL-TOV to pre-compensate the polarization dependent reflection of the optical elements between the sample and the generation point, in this case the diffraction grating. To optimize and verify the circular polarization at the sample plane, we use an extreme-UV polarization analyzer based on nanoscale slit arrangement that allows for *in-situ* polarization measurement in a single acquisition [62].

**Imaging scheme.**
We employ holographically guided lensless imaging, in which the sample mask contains a circular FOV aperture, two or more holographic reference holes, and an array of auxiliary holes. The masks are designed according to the following criteria:

1. A confinement of the exit wave to fulfill the sampling requirement for phase retrieval [58]. This criterion defines the maximum size of the FOV aperture, which in our case was successfully tested for circular apertures with diameters of 2, 3, 4, 5 and 6 μm.

2. Sufficiently small holographic reference holes (diameters between 200 nm and 400 nm) to provide for exact information on the sample geometry and the real-space support via FTH directly from the measured data. Having two reference holes makes the automatic support deconvolution process more straightforward.

3. An array of additional auxiliary holes of a specific shape and arrangement is introduced in the vicinity of the sample to ensure homogeneous coverage of the detector with a strong scattering signal that interferes with weak magneto-optical signal from the FOV aperture and coherently enhances it above the detector noise level. This reduces the required dose by more than an order of magnitude, depending on required resolution.

For every delay frame in the pump-probe movie, we record two sets of diffraction data using a left- and right-handed circularly polarized harmonic illumination. Each data set is composed of 10-20 diffraction patterns with 3 to 30 seconds exposure time (depending on the mask geometry and NA used) to increase

SNR and the dynamic range of the diffraction data. In the case of high-resolution imaging experiments shown in Fig. 3, a total exposure time of up to 15 minutes was required for sub-wavelength spatial resolution reconstructions. To further improve the SNR for these data sets, we acquire short and long exposure time diffraction patterns that are subsequently merged into a single HDR data set. Starting from a random first guess, deconvolved holographic support and using a modified RAAR algorithm [63], we first reconstruct data set with left-handed circular polarized probe using 130-150 iterations and afterwards reconstruct the data set of the opposite helicity. If both data sets are accurately reconstructed and precisely aligned, the ratio of these two independently reconstructed exit wave amplitudes of opposite helicities eliminate non-dichroic contributions, providing for a pure magnetic phase and absorption contrast image. To accelerate the iterative process to just 10 iterations and to skip the sub-pixel alignment procedure, the reconstruction for the opposite helicity (as well as for the all data sets of a given time series) can be initiated from a previously obtained reconstruction. Importantly, such an approach mitigates possible phase retrieval and imaging artefacts for diffraction data with poor SNR, low oversampling ratio, or insufficient coherence properties of the probe. Irrespective of the initial random guess, the dichroic reconstructions are virtually identical (as evident form the PRTF plotted in Fig. 2c), thus no averaging of several reconstructions or filtering is required for diffraction data with high SNR. Furthermore, no additional phase retrieval constraints have to be imposed, and we attribute this to the improved spatial and temporal coherence of the developed HHG source as well as the coherent signal enhancement mechanism provided by the auxiliary reference holes.

We note that time-invariant regions of the sample, i.e., the auxiliary and reference holes, can be enforced as an additional real-space constraint in the reconstruction process to potentially relax the SNR requirement for each delay frame, following alternative approaches specifically designed for pump-probe diffraction data [64].

**Resolution estimate.**
To estimate the spatial resolution, we first measure the accuracy of the phase retrieval process in the far-field using a well-established PRTF method [65]. Figure 3c shows the PRTF plot for 20 individual reconstructions (for the data with NA=0.65) initiated from a random first guess demonstrating a consistent phase retrieval throughout the entire recorded far-field as expected from a diffraction limited imaging system. Additionally, we verify the spatial resolution in real space as the smallest resolvable feature in the reconstruction. We find multiple localized domains, bubbles (and possibly skyrmions) with sub-50 nm transverse sizes that are clearly resolved. The smallest dimensions in the investigated samples are magnetic domain walls that are expected to be in the range between 10 and 20 nm [66]. To estimate the sharpness of the reconstructed domain walls we used an improved knife-edge technique that measures the average transition sharpness between all identified magnetic domain boundaries throughout the image. In this way, we obtain a conservative estimate and the characteristic resolution, avoiding artificially sharp transitions from a single lineout in the presence of noise. Furthermore, to rule out any influence of the segmentation algorithm in locating domain walls, we identified the domain wall positions in one data set, and measured the final resolution using an independent data set of the same structure. The averaged transition in contrast occurs over less than two pixels, consistent with the expected physical domain wall sharpness and the 16 nm spatial resolution provided by the PRTF. The reconstructed image also exhibits multiple domain walls appearing as a single-pixel step.

The temporal resolution of our approach is conservatively estimated as 40 fs using an optical cross-correlation of the 35 fs pump beam (verified with SPIDER) and expected sub-10 fs harmonic probe. First, we measure the spatio-temporal shear introduced by the diffraction grating to the probe pulse. Second, we compute a fraction of this shear incident onto a 3µm large FOV of the object. Finally, we consider the non-collinear excitation on the FOV and measure the convolution of the resulting pulses.

**Magnetic film fabrication**

The magnetic Co/Pd ML films have the following structure: [Pd(0.75 nm)/Co(0.55 nm)]$_9$. The MLs were prepared at RT by dc magnetron sputtering from elemental targets on Silicon and Si3N4 membranes. The sputter process was carried out using an Ar working pressure of $3.5 \times 10^{-3}$ mbar in an ultra-high vacuum chamber (base pressure $<10^{-8}$ mbar). For all films, 1.5 nm of Cr and 2.0 nm of Pd were used as seed layer. To protect the films from corrosion 2 nm of Pd were used as cover layer. The thicknesses of the layers were estimated from the areal densities measured by a quartz balance during deposition.

**Acknowledgments**

We gratefully acknowledge the support and insightful discussions with Tim Salditt, Marcell Möller, Thomas Danz, Tobias Heinrich, Philipp Buchsteiner.

O.K. gratefully acknowledges funding from the European Union's Horizon 2020 research and innovation programme under the Marie Skłodowska-Curie grant agreement No.752533. This work was funded by the Deutsche Forschungsgemeinschaft (DFG) in the Collaborative Research Center "Nanoscale Photonic Imaging" (DFG-SFB 755, project C08). S.Z acknowledges funding from the Campus Laboratory for Advanced Imaging, Microscopy and Spectroscopy (AIMS) at the University of Göttingen.

**Competing interests:** The authors declare no competing interests.

**Data availability:** The data of this study is available from the corresponding author upon request.

**Author contributions:** S.Z., O.K. and C.R. conceived and designed the experiment with contributions from M.L. and M.S. The samples were designed by S.Z. O.K., M.S., M.H., and M.A., and fabricated by M.H., M.S. and M.A. Measurements and data analysis were performed by S.Z. and O.K. The manuscript was written by S.Z., O.K and C.R. with contributions from all authors.